\documentclass[default]{aastex701}

\shortauthors{Asai et al.}
\accepted{\today}
\submitjournal{ApJL}
\begin{document}

\title{Chromospheric Dynamics of an Umbral Flare Kernel 
-- Based on Coordinated SUNRISE~III SCIP and Domeless Solar Telescope Observations}

\author[orcid=0000-0002-5279-686X,gname=Ayumi, sname=Asai]{Ayumi Asai}
\affiliation{Astronomical Observatory, Kyoto University, Kitashirakawa-Oiwakecho, Sakyo-ku, Kyoto, 606-8502, Japan}
\email[show]{asai@kwasan.kyoto-u.ac.jp}  

\author[orcid=0000-0002-2165-7998]{Satoru UeNo} 
\affiliation{Astronomical Observatory, Kyoto University, Kamitakara, Takayama, Gifu, 506-1314, Japan}
\email{ueno@kwasan.kyoto-u.ac.jp}

\author[orcid=0000-0002-1043-9944,sname='Matsumoto']{Takuma~Matsumoto} 
\affiliation{Centre for Integrated Data Science, Institute for Space-Earth Environmental Research, Nagoya University, Furocho, Chikusa-ku, Nagoya, Aichi 464-8601, Japan}
\email{takuma.matsumoto@gmail.com}

\author[orcid=0000-0002-7044-6281,sname='Oba']{Takayoshi~Oba} 
\affiliation{Advanced Research Center for Space Science and Technology, Institute of Science and Engineering, Kanazawa University, Kakuma-machi, Kanazawa, Ishikawa 920-1192, Japan}
\affiliation{Max-Planck-Institut f\"{u}r Sonnensystemforschung, Justus-von-Liebig-Weg 3, 37077 G\"{o}ttingen, Germany}
\email{oba@mps.mpg.de}

\author[orcid=0000-0002-5054-8782,sname='Katsukawa']{Yukio~Katsukawa} 
\affiliation{National Astronomical Observatory of Japan, 2-21-1 Osawa, Mitaka, Tokyo 181-8588, Japan}
\affiliation{Department of Earth and Planetary Science, The University of Tokyo, 7-3-1, Hongo, Bunkyo-ku, Tokyo 113-0033, Japan}
\affiliation{Department of Astronomical Science, The Graduate University for Advanced Studies (SOKENDAI), 2-21-1 Osawa, Mitaka, Tokyo 181-8588, Japan}
\email{yukio.katsukawa@nao.ac.jp}

\author[orcid=0000-0001-5616-2808,sname='Kubo']{Masahito~Kubo} 
\affiliation{National Astronomical Observatory of Japan, 2-21-1 Osawa, Mitaka, Tokyo 181-8588, Japan}
\email{masahito.kubo@nao.ac.jp}

\author[orcid=0000-0002-4669-5376,sname='Ishikawa']{Ryohtaroh~T.~Ishikawa} 
\affiliation{National Institute for Fusion Science, 322-6 Oroshi-cho, Toki City 509-5292, Japan}
\email{ishikawa.ryohtaro@nifs.ac.jp}

\author[orcid=0000-0001-7452-0656,sname='Kawabata']{Yusuke~Kawabata} 
\affiliation{National Astronomical Observatory of Japan, 2-21-1 Osawa, Mitaka, Tokyo 181-8588, Japan}
\email{kawabata.yusuke@nao.ac.jp}

\author[orcid=0000-0001-5686-3081,sname='Hara']{Hirohisa~Hara} 
\affiliation{National Astronomical Observatory of Japan, 2-21-1 Osawa, Mitaka, Tokyo 181-8588, Japan}
\affiliation{Department of Astronomical Science, The Graduate University for Advanced Studies (SOKENDAI), 2-21-1 Osawa, Mitaka, Tokyo 181-8588, Japan}
\affiliation{Institute of Space and Astronautical Science, Japan Aerospace Exploration Agency, 3-1-1, Yoshinodai, Chuo-ku, Sagamihara, Kanagawa 252-5210, Japan}
\email{hirohisa.hara@nao.ac.jp}

\author[orcid=0000-0001-6793-8528,sname='Naito']{Yoshihiro~Naito} 
\affiliation{Department of Astronomical Science, The Graduate University for Advanced Studies (SOKENDAI), 2-21-1 Osawa, Mitaka, Tokyo 181-8588, Japan}
\affiliation{National Astronomical Observatory of Japan, 2-21-1 Osawa, Mitaka, Tokyo 181-8588, Japan}
\email{yoshihiro.naito@grad.nao.ac.jp}

\author[orcid=0000-0003-4764-6856,sname='Shimizu']{Toshifumi~Shimizu} 
\affiliation{Department of Earth and Planetary Science, The University of Tokyo, 7-3-1, Hongo, Bunkyo-ku, Tokyo 113-0033, Japan}
\affiliation{Institute of Space and Astronautical Science, Japan Aerospace Exploration Agency, 3-1-1, Yoshinodai, Chuo-ku, Sagamihara, Kanagawa 252-5210, Japan}
\email{shimizu.toshifumi@isas.jaxa.jp}

\author[orcid=0000-0002-3418-8449,sname='Solanki']{Sami~K.~Solanki} 
\affiliation{Max-Planck-Institut f\"{u}r Sonnensystemforschung, Justus-von-Liebig-Weg 3, 37077 G\"{o}ttingen, Germany}
\email{solanki@mps.mpg.de}

\author[orcid=0000-0003-1459-7074,sname='Lagg']{Andreas~Lagg} 
\affiliation{Max-Planck-Institut f\"{u}r Sonnensystemforschung, Justus-von-Liebig-Weg 3, 37077 G\"{o}ttingen, Germany}
\email{lagg@mps.mpg.de}

\author[orcid=0000-0002-9972-9840,sname='Gandorfer']{Achim~Gandorfer} 
\affiliation{Max-Planck-Institut f\"{u}r Sonnensystemforschung, Justus-von-Liebig-Weg 3, 37077 G\"{o}ttingen, Germany}
\email{gandorfer@mps.mpg.de}

\author[orcid=0000-0002-3387-026X,sname='del~Toro~Iniesta']{Jose~Carlos~del~Toro~Iniesta} 
\affiliation{Instituto de Astrof\'{i}sica de Andaluc\'{i}a, CSIC, Glorieta de la Astronom\'{i}a s/n, 18008 Granada, Spain}
\affiliation{Spanish Space Solar Physics Consortium}
\email{jti@iaa.es}

\author[orcid=0000-0002-0787-8954,sname='Bernasconi']{Pietro~Bernasconi} 
\affiliation{Johns Hopkins University Applied Physics Laboratory, 11100 Johns Hopkins Road, Laurel, Maryland, USA}
\email{pietro.bernasconi@jhuapl.edu}

\author[sname='Berkefeld']{Thomas~Berkefeld} 
\affiliation{Institut f\"{u}r Sonnenphysik (KIS), Georges-K\"{o}hler-Allee 401a, 79110 Freiburg, Germany}
\email{thomas.berkefeld@leibniz-kis.de}

\author[orcid=0009-0009-4425-599X,sname='Feller']{Alex~Feller} 
\affiliation{Max-Planck-Institut f\"{u}r Sonnensystemforschung, Justus-von-Liebig-Weg 3, 37077 G\"{o}ttingen, Germany}
\email{feller@mps.mpg.de}

\author[orcid=0000-0001-6317-4380,sname='Riethm\"{u}ller']{Tino~L.~Riethm\"{u}ller} 
\affiliation{Max-Planck-Institut f\"{u}r Sonnensystemforschung, Justus-von-Liebig-Weg 3, 37077 G\"{o}ttingen, Germany}
\email{riethmueller@mps.mpg.de}

\author[orcid=0000-0001-9228-3412,sname='\'{A}lvarez-Herrero']{Alberto~\'{A}lvarez-Herrero} 
\affiliation{Instituto Nacional de T\'ecnica Aeroespacial (INTA), Ctra. de Ajalvir, km. 4, E-28850 Torrej\'{o}n de Ardoz, Spain}
\affiliation{Spanish Space Solar Physics Consortium}
\email{alvareza@inta.es}

\author[orcid=0000-0003-3490-6532,sname='Smitha']{H.~N.~Smitha} 
\affiliation{Max-Planck-Institut f\"{u}r Sonnensystemforschung, Justus-von-Liebig-Weg 3, 37077 G\"{o}ttingen, Germany}
\email{narayanamurthy@mps.mpg.de}

\author[orcid=0000-0001-8829-1938,sname='Orozco~Su\'{a}rez']{David~Orozco~Su\'{a}rez} 
\affiliation{Instituto de Astrof\'{i}sica de Andaluc\'{i}a, CSIC, Glorieta de la Astronom\'{i}a s/n, 18008 Granada, Spain}
\affiliation{Spanish Space Solar Physics Consortium}
\email{orozco@iaa.es}

\author[sname='Grauf']{Bianca~Grauf} 
\affiliation{Max-Planck-Institut f\"{u}r Sonnensystemforschung, Justus-von-Liebig-Weg 3, 37077 G\"{o}ttingen, Germany}
\email{grauf@mps.mpg.de}

\author[sname='Carpenter']{Michael~Carpenter} 
\affiliation{Johns Hopkins University Applied Physics Laboratory, 11100 Johns Hopkins Road, Laurel, Maryland, USA}
\email{michael.carpenter@jhuapl.edu}

\author[sname='Bell']{Alexander~Bell} 
\affiliation{Institut f\"{u}r Sonnenphysik (KIS), Georges-K\"{o}hler-Allee 401a, 79110 Freiburg, Germany}
\email{albe@leibniz-kis.de}

\author[orcid=0000-0001-7764-6895,sname='Mart\'{i}nez~Pillet']{Valentín~Mart\'{i}nez~Pillet} 
\affiliation{Instituto de Astrof\'{i}sica de Canarias, V\'{i}a L\'{a}ctea, s/n, E-38205 La Laguna, Spain}
\affiliation{Spanish Space Solar Physics Consortium}
\email{vmpillet@iac.es}

\author[orcid=0000-0001-7696-8665,sname='Gizon']{Laurent~Gizon} 
\affiliation{Max-Planck-Institut f\"{u}r Sonnensystemforschung, Justus-von-Liebig-Weg 3, 37077 G\"{o}ttingen, Germany}
\affiliation{Institut f\"{u}r Astrophysik und Geophysik, Georg-August-Universit\"{a}t G\"{o}ttingen, 37077 G\"{o}ttingen, Germany}
\email{gizon@mps.mpg.de}

\author[orcid=0000-0002-7318-3536,sname='Bail\'{e}n']{Francisco~Javier~Bail\'{e}n} 
\affiliation{Instituto de Astrof\'{i}sica de Andaluc\'{i}a, CSIC, Glorieta de la Astronom\'{i}a s/n, 18008 Granada, Spain}
\affiliation{Spanish Space Solar Physics Consortium}
\email{fbailen@iaa.es}

\author[orcid=0000-0002-2055-441X,sname='Blanco~Rodr\'{i}guez']{Julian~Blanco~Rodr\'{i}guez} 
\affiliation{Universitat de Valencia Catedr\'{a}tico Jos\'{e} Beltr\'{a}n 2, E-46980 Paterna-Valencia, Spain}
\affiliation{Spanish Space Solar Physics Consortium}
\email{julian.blanco@uv.es}

\author[orcid=0000-0003-4319-2009,sname='Castellanos~Dur\'{a}n']{Juan~Sebasti\'{a}n~Castellanos~Dur\'{a}n} 
\affiliation{Max-Planck-Institut f\"{u}r Sonnensystemforschung, Justus-von-Liebig-Weg 3, 37077 G\"{o}ttingen, Germany}
\email{castellanos@mps.mpg.de}

\author[orcid=0009-0002-6808-5154,sname='Harnes']{Edvarda~Harnes} 
\affiliation{Max-Planck-Institut f\"{u}r Sonnensystemforschung, Justus-von-Liebig-Weg 3, 37077 G\"{o}ttingen, Germany}
\email{harnes@mps.mpg.de}

\author[orcid=0000-0001-6029-7529,sname='Hoelken']{Johannes~Hoelken} 
\affiliation{Max-Planck-Institut f\"{u}r Sonnensystemforschung, Justus-von-Liebig-Weg 3, 37077 G\"{o}ttingen, Germany}
\email{hoelken@mps.mpg.de}

\author[orcid=0000-0003-1409-1145,sname='Iglesias']{Francisco~A.~Iglesias} 
\affiliation{Max-Planck-Institut f\"{u}r Sonnensystemforschung, Justus-von-Liebig-Weg 3, 37077 G\"{o}ttingen, Germany}
\affiliation{Grupo de Estudios en Heliof\'{i}sica de Mendoza, CONICET, Universidad de Mendoza, Boulogne sur Mer 683, 5500 Mendoza, Argentina}
\email{iglesias@mps.mpg.de}

\author[orcid=0000-0003-0175-6232,sname='Siu-Tapia']{Azaymi~L.~Siu-Tapia} 
\affiliation{Instituto de Astrof\'{i}sica de Andaluc\'{i}a, CSIC, Glorieta de la Astronom\'{i}a s/n, 18008 Granada, Spain}
\affiliation{Spanish Space Solar Physics Consortium}
\email{siu@iaa.es}

\author[orcid=0000-0003-1483-4535,sname='Strecker']{Hanna~Strecker} 
\affiliation{Instituto de Astrof\'{i}sica de Andaluc\'{i}a, CSIC, Glorieta de la Astronom\'{i}a s/n, 18008 Granada, Spain}
\affiliation{Spanish Space Solar Physics Consortium}
\email{streckerh@iaa.es}

\author[orcid=0000-0003-1971-5551,sname='Vukadinovi\'{c}']{Du\v{s}an~Vukadinovi\'{c}} 
\affiliation{Institut f\"{u}r Physik, Universit\"{a}t Graz, Universit\"{a}tsplatz 5, 8010 Graz, Austria}
\affiliation{Max-Planck-Institut f\"{u}r Sonnensystemforschung, Justus-von-Liebig-Weg 3, 37077 G\"{o}ttingen, Germany}
\email{vukadinovic@mps.mpg.de}


\collaboration{all}{\textsc{Sunrise~iii} team}

\begin{abstract}

We report imaging spectroscopic observations of an M1.4 solar flare obtained during a coordinated observation between the infrared spectropolarimeter SCIP onboard the \textsc{Sunrise~iii} balloon mission and Domeless Solar Telescope (DST) at Hida Observatory, Kyoto University. 
The flare that occurred on 2024 July 13 in NOAA Active Region 13738 exhibited a compact flare kernel located within a sunspot umbra.
SCIP performed rapid slit-scan observations over a field of view of $58^{\prime\prime} \times 58^{\prime\prime}$ around the umbra with a cadence of 40~s, covering infrared chromospheric and upper-photospheric lines including Ca~{\sc ii} 8498/8542~{\AA} and K~{\sc i} D$_1$. 
At the same time, DST observed a wider surrounding region with a cadence of 25~s in H$\alpha$, Ca~{\sc ii} 8542~{\AA}, and Na~{\sc i} D$_1$/D$_2$. 
Clear flare-related brightenings are detected in all chromospheric lines observed by SCIP and DST, while no significant enhancement is found in photospheric lines. 
The high spatial resolution of SCIP reveals fine substructures within the kernel on spatial scales of order 1000~km, which appear smeared in ground-based observations.
The spectral profiles exhibit temporally and spatially varying Doppler shifts and line broadenings, indicating complex, fine-scale plasma motions in the chromosphere. 
These results suggest that the observed red asymmetry arises from the temporal succession of multiple fine-scale kernels, as revealed by SCIP, rather than from a single continuous process.

\end{abstract}


\section{Introduction} 

Solar flares are explosive energy release events in the solar atmosphere.
Magnetic energy stored in the corona is rapidly converted via magnetic reconnection into thermal energy, bulk plasma motions, and nonthermal particles \citep[e.g.,][]{Shibata2011}.
The released energy is transported from the corona into the chromosphere, driving a rapid and highly dynamic response of the chromospheric plasma.
This energy deposition produces characteristic chromospheric phenomena, including flare ribbons with compact flare kernels inside, which are one of the most dominant observational signatures of solar flares.
The so-called CSHKP magnetic reconnection model for solar flares \citep{Carmichael1964,Sturrock1966,Hirayama1974,KoppPneuman1976}, which has been accepted as a standard model, suggests that successive reconnection causes the apparent separation of flare ribbons \citep{Sturrock1992,Svestka1992}.

Energy transport by thermal conduction and/or nonthermal particles into the chromosphere causes characteristic flare brightenings, spectral distortions, and Doppler shifts in chromospheric lines.
Intensity enhancements dominant in the red wings are known as “red asymmetry” of the lines \citep{Svestka1976}.
Spectroscopic observations of flare kernels have demonstrated that this red asymmetry corresponds to downward motions with velocities of several tens of km~s~$^{-1}$, which have been interpreted as chromospheric condensation \citep{Ichimoto1984,Falchi1997}.
Multi-line spectroscopic observations further revealed that the magnitude of the red asymmetry depends on the spectral line.
\citet{Shoji1995} showed that strong red-shifted components appear prominently in H$\alpha$ and Ca~{\sc ii} K lines, whereas Na~{\sc i} D$_1$/D$_2$ lines exhibit only weak red shifts of a few km~s~$^{-1}$.
They interpreted these findings to imply that stronger downward motions occur preferentially in the upper chromosphere, leading to a formation-height dependence of chromospheric condensation. 
A “blue asymmetry” has also been reported to appear at very early phases of flares \citep{Svestka1976} or in higher-formation chromospheric lines such as  Mg~{\sc ii}~h \citep{Tei2018}.
This is probably caused by upflows of cool plasma ($\sim 10^{4}$~K) lifted by expanding hot plasma driven by deep penetration of nonthermal electrons.

The spatial distribution of the red asymmetry provides important clues to where strong energy deposition occurs within flare ribbons.
Early imaging observations suggested that the red asymmetry is widely distributed along flare ribbons \citep[e.g.,][]{Janssens1970,Svestka1976,Tang1983}, while subsequent studies have demonstrated that the strongest red-shifted components are preferentially localized near the outer edges of flare ribbons, where newly reconnected magnetic loops are rooted \citep{Falchi1997}.
\citet{Asai2012} showed that the pronounced red asymmetry is confined to narrow regions with widths of order 1000 -- 2000~km at the outermost edges of flare ribbons, by using narrowband H$\alpha$ filtergram observations.
They further reported that the enhancement of red asymmetry often precedes the intensity increase at individual kernels, suggesting that chromospheric condensation develops rapidly at sites of newly established energy deposition.

Reproducing the observed spectral distortions and Doppler shifts has long been a theoretical and numerical challenge owing to the highly nonlinear and dynamic nature of the chromospheric plasma \citep[e.g.,][]{Fisher1985a,Fisher1985b}.
Following these pioneering works, recent radiative hydrodynamic simulations have significantly advanced our understanding of chromospheric responses to flare energy input \citep[e.g.,][]{Allred2015,Kowalski2017,Kowalski2022}.
These models predict complex, height-dependent responses of chromospheric spectral lines to nonthermal electron beams and thermal conduction, and provide an essential theoretical framework for interpreting multi-line spectroscopic observations of flare kernels.

A combination of high-cadence imaging and spectroscopy has begun to probe the fine-scale temporal and spatial evolution of flare ribbons and kernels.
Observations with Interface Region Imaging Spectrograph \citep[IRIS;][]{DePontieu2014IRIS} have revealed rapidly evolving chromospheric flows and transient red-shifted components in flare ribbons and kernels \citep{Graham2015,Graham2020}.
\citet{Ashfield2026} used coordinated IRIS and the Swedish 1~m Solar Telescope (SST) observations to connect chromospheric condensation velocities with quasi-periodic energy deposition.
Recent high-spatial resolution observations with the Daniel K. Inouye Solar Telescope \citep[DKIST;][]{Rimmele2020DKIST} and \textsc{Sunrise~iii} balloon mission \citep{solanki2026SpecialIssues,Korpi-Lagg2025SUNRISE3,Bartho2011SUNRISE,Sokanki2010,Solanki2017} have revealed fine-scale structures in flare ribbons and kernels in the chromosphere and photosphere \citep{Yadav2024,Tamburri2025,Tamburri2026,Chitta2026}.
These results are part of a broader observational picture in which flare ribbons are increasingly recognized as being composed of fine substructures.
For example, \citet{Jing2016GST} reported sub-arcsecond structures in flare ribbons by using the Goode Solar Telescope at Big Bear Solar Observatory, while SST observations have identified elongated small-scale features (“riblets”) within flare ribbons \citep{ThoenFaber2026,Singh2026}.
The Extreme Ultraviolet Imager \citep[EUI;][]{Rochus2020} onboard Solar Orbiter has also revealed extreme ultraviolet counterparts of such fine structures \citep{Chitta2026EUI,Sun2026}.
However, simultaneous high-spatial-resolution, multi-line chromospheric spectroscopy that connects fine-scale kernel morphology with line-profile asymmetries remains limited, especially for flare kernels in sunspot umbrae.

In this paper, we report coordinated imaging spectroscopic observations of an M1.4 solar flare that occurred on 2024 July 13 in NOAA Active Region 13738, obtained with the \textsc{Sunrise} Chromospheric Infrared spectro-Polarimeter \citep[SCIP;][]{Katsukawa2026SCIP} onboard the \textsc{Sunrise~iii} and with Domeless Solar Telescope \citep[DST;][]{Nakai1985DST} at Hida Observatory, Kyoto University.
By combining these observations, we investigate the temporal evolution, spatial structure, and spectral characteristics of the umbral flare kernel, and discuss their implications for chromospheric condensation and energy deposition in solar flares.

\section{Observations} 

\begin{figure*}[t]
\centering
\includegraphics[width=14.4cm]{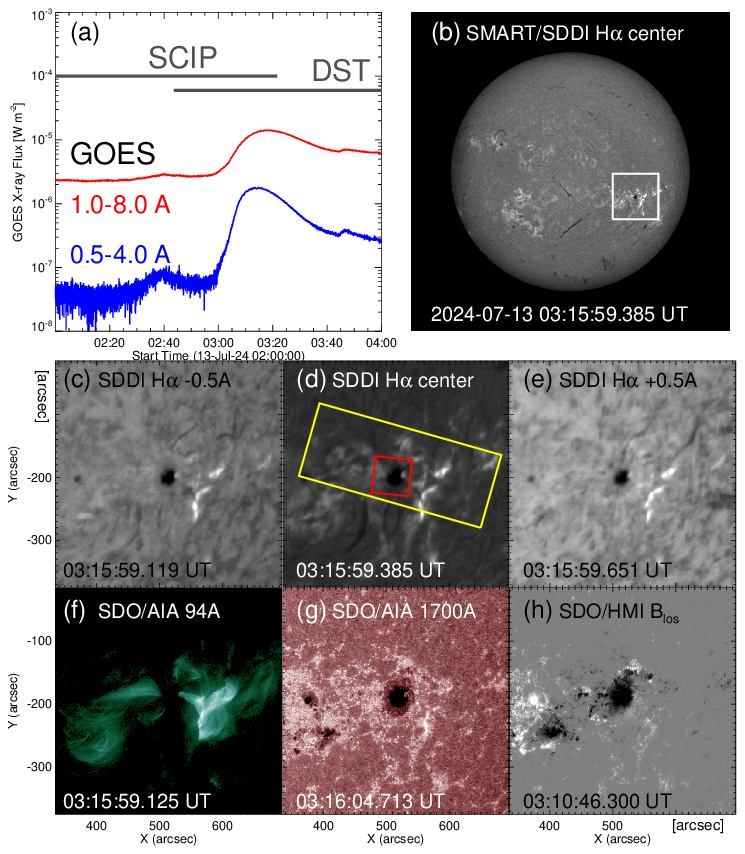}
\caption{(a) The GOES soft X-ray light curves in 1.0 -- 8.0~{\AA} and 0.5 -- 4.0~{\AA} from 02:00~UT to 04:00~UT on 2024 July 13 are shown as red and blue lines, respectively.
The gray horizontal lines indicate the time intervals of the SCIP and DST observations.
(b) The H$\alpha$ line center full-disk image taken by SMART/SDDI at 03:16~UT on 2024 July 13.
Solar north and west are at the top and right.
The white box indicates the target active region NOAA 13738, and corresponds to the field of view in Figures~\ref{fig:ltcv}(c) -- (h).
(c), (d), and (e) The SMART/SDDI images in H$\alpha$ $-0.5$~{\AA}, center, and $+0.5$~{\AA}, respectively. 
The yellow and red boxes are the field of views of DST and SCIP, respectively.
(f) and (g) SDO/AIA images in 94~{\AA} and 1700~{\AA}.
(h) line-of-sight magnetogram taken at 03:10:46~UT on 2024 July 13 with SDO/HMI.}
\label{fig:ltcv}
\end{figure*}

We observed an M1.4 solar flare, which occurred on 2024 July 13 within the Active Region (AR) NOAA 13738 (S09$^{\circ}$, W37$^{\circ}$) during the coordinated observation between \textsc{Sunrise~iii}/SCIP and DST.
The flare started at 02:55~UT and peaked at 03:18~UT.
Figure~\ref{fig:ltcv}(a) shows the light curves of the soft X-rays in 1.0 -- 8.0~{\AA} (red) and 0.5 -- 4.0~{\AA} (blue) obtained with GOES.
The horizontal, gray, thick lines indicate the time intervals for the SCIP and DST observations.
The coordinated observations with DST and SCIP began before the flare onset and continued through the flare peak, allowing us to capture the temporal evolution of the flare kernel with simultaneous spectroscopic measurements.
Figure~\ref{fig:ltcv}(b) shows the H$\alpha$ full-disk image obtained at 03:16~UT by Solar Dynamics Doppler Imager \citep[SDDI;][]{Ichimoto2017SDDI} mounted on Solar Magnetic Activity Research Telescope \citep[SMART;][]{Ueno2004SMART} at Hida Observatory, Kyoto University.
SMART/SDDI regularly observes the full-disk Sun at 73 wavelength points between ${\pm}$9.0~{\AA} from the H$\alpha$ line center with a spectral sampling of 0.25~{\AA}.
The cadence of one complete spectral scan is 12 seconds.
The pixel size is 1.23$^{\prime\prime}$.
The white box indicates the region of the panels (c) -- (h).

Figures~\ref{fig:ltcv}(c), (d), and (e) show close-up H$\alpha$ images of the AR taken at 03:16~UT in H$\alpha$ $-0.5$~{\AA}, center, and $+0.5$~{\AA}, respectively.
In these H$\alpha$ images, a compact kernel in the umbra and an extended ribbon in the western side of the umbra are seen.
Figures~\ref{fig:ltcv}(f) and (g) show AIA 94~{\AA} and 1700~{\AA} images obtained with 
the Atmospheric Imaging Assembly \citep[AIA;][]{Lemen2012AIA} on board the Solar Dynamics Observatory \citep[SDO;][]{Pesnell2012SDO}.
The AIA 1700~{\AA} image clearly shows the flare ribbon, while the flare kernel in the umbra is, although recognizable, much fainter.
The flare loops are visible in the AIA 94~{\AA} image.
Although its structure is complex, the umbral flare kernel is primarily connected to the flare ribbon on the southwest side.
Figure~\ref{fig:ltcv}(h) is a line-of-sight magnetogram obtained by the Helioseismic and Magnetic Imager \citep[HMI;][]{Scherrer2012HMI} on board SDO.
The umbra, which shows the compact flare kernel, has negative magnetic polarity.

SCIP observed the target region from 01:34:16~UT to 03:21:35~UT.
SCIP ran in ''rapid mode'' during the flare, which means that only the Stokes-$I$ information was obtained.
The size of the FOV of 58$^{\prime\prime}$ $\times$ 58$^{\prime\prime}$ is marked by the red box in Figure~\ref{fig:ltcv}(d).
The spatial sampling along the slit is 0.094$^{\prime\prime}$, and the raster step size is 0.094$^{\prime\prime}$.
The raster cadence is about 40 seconds.
Although the full Stokes parameters were not obtained with the rapid mode, the rotating waveplate was in operation, which modulated the intensity (Stokes-$I$ signal).
Therefore, to remove the intensity modulation, we binned every four consecutive steps, corresponding to half a rotation of the waveplate.
This resulted in a spatial sampling of 0.37$^{\prime\prime}$ in the scan direction, and in a step time of 0.256 seconds in the scan.

During the observation, the solar elevation as seen from \textsc{Sunrise~iii} was about $-3^{\circ}$, i.e., the Sun was slightly below the local horizon, so that the line of sight passed through a part of Earth’s atmosphere.
This caused image distortions and reduced the effective spatial resolution.
We estimate that the effective spatial resolution during the flare was approximately 0.4$^{\prime\prime}$ -- 1.0$^{\prime\prime}$, while the nominal spatial resolution of SCIP is 0.21$^{\prime\prime}$, corresponding to the diffraction limit at a wavelength of 8500~{\AA} \citep{Katsukawa2026SCIP}.
The observed wavelength ranges are recorded in two spectral channels around 8500~{\AA} and 7700~{\AA} \citep{solanki2026SpecialIssues,Katsukawa2026SCIP}.
These include the chromospheric Ca~{\sc ii} 8498.0~{\AA} and 8542.1~{\AA} (hereafter, Ca~8498/Ca~8542) lines, and K~{\sc i}~D$_1$ (7699.0~{\AA}; hereafter K~D$_1$) line as well as several photospheric lines such as Fe~{\sc i} 8514.1~{\AA}.
The spectral sampling is 39.5~m{\AA} for the 8500~{\AA} channel and 36.0~m{\AA} for the 7700~{\AA} channel.
The instrumental spectral resolution estimated from the measured spectral point spread function (PSF) is 37.9~m{\AA} for the 8500~{\AA} channel and 32.0~m{\AA} for the 7700~{\AA} channel.

DST performed slit-scan spectroscopic observations from 02:43:34~UT to 04:01:03~UT with the wavelength windows of H$\alpha$ (6562.8~{\AA}), Na~{\sc i}~D$_1$/D$_2$ (5895.9/5889.9~{\AA}; hereafter, Na~D$_1$/D$_2$), and Ca~8542. 
The bottom gray horizontal line in Figure~\ref{fig:ltcv}(a) indicates the time range.
The spatial sampling along the slit is 0.38$^{\prime\prime}$, and the raster step size is 0.68$^{\prime\prime}$.
The effective spatial resolution was limited by seeing and is estimated to be approximately 0.7$^{\prime\prime}$ -- 2$^{\prime\prime}$ during the observations.
The spectral sampling is about 17.3~m{\AA} for the Ca~8542 line.
The spectral resolution is estimated to be 40.1~m{\AA} from the spectrograph setting and the slit width.
The raster cadence is about 25 seconds.
The observations were occasionally affected by passing clouds.
Because both SCIP and DST data were obtained by raster scans, the observation times indicated for raster images and spectra in this paper refer to the start times of the corresponding raster scans, unless otherwise noted. 

Based on previous formation-height estimates \citep[e.g.,][]{Vernazza1981VAL,Carlsson2016,Lagg2017,Katsukawa2026SCIP}, these line-core images sample different atmospheric layers under quiet-sun or standard atmospheric conditions.
The Fe~{\sc i} lines are mainly photospheric. 
The K~D$_1$ and the Na~D$_1$/D$_2$ lines sample the photosphere--chromosphere interface, with Na~D$_1$/D$_2$ generally sampling slightly higher layers than K~D$_1$.
The H$\alpha$ and Ca~8498/Ca~8542 lines are chromospheric.
However, these formation heights are model-dependent and can change substantially in a sunspot umbra, especially under flare conditions.

\section{Fine Structures of Umbral Flare Kernel} 

\begin{figure*}[t]
\centering
\includegraphics[width=14.4cm]{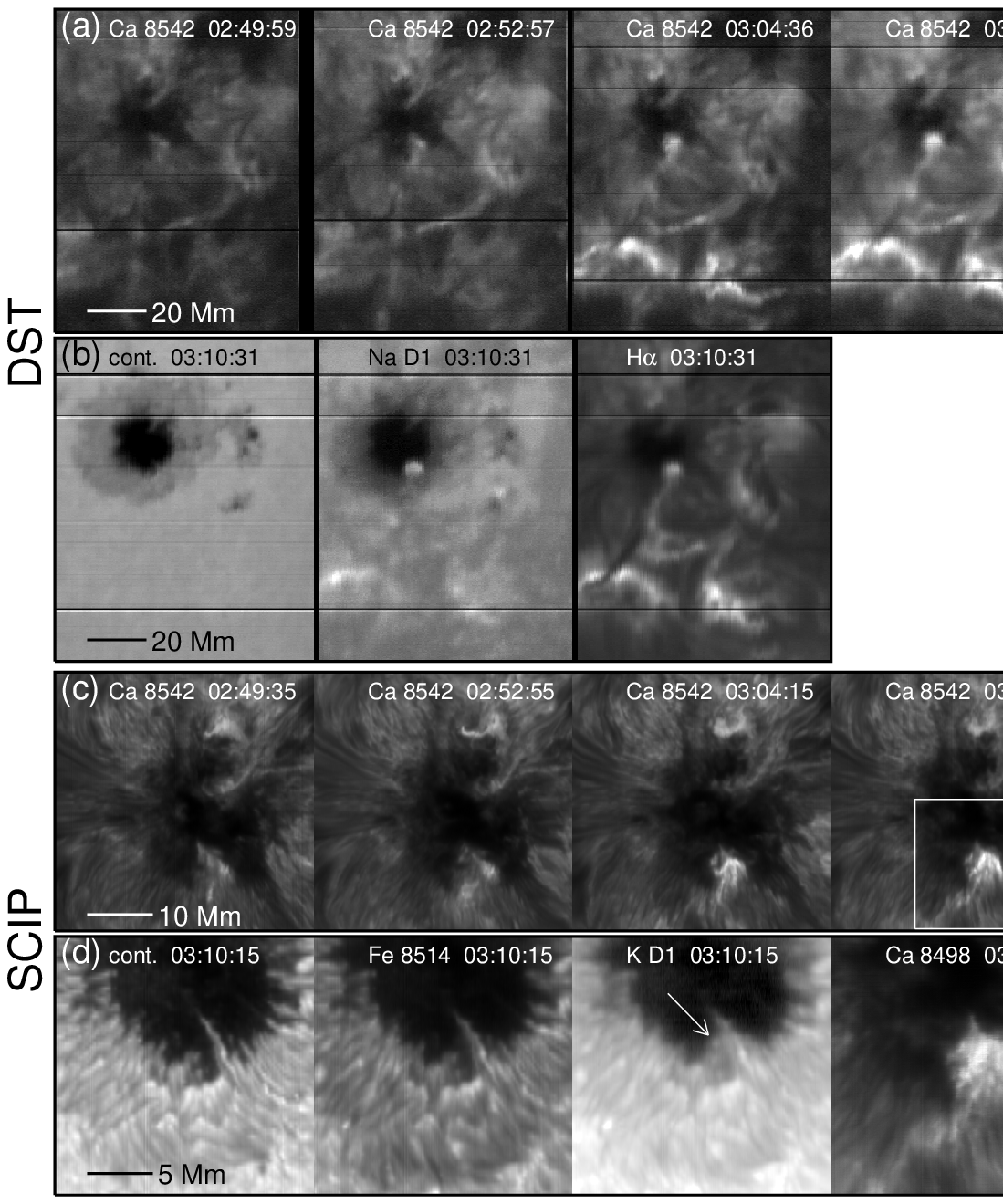}
\caption{(a) Time sequence images of the flare kernel observed with DST in Ca~8542 line center.
(b) DST images obtained at 03:10:31~UT in different spectral lines.
From left to right, continuum, Na~D$_1$ line center, and H$\alpha$ line center are shown.
Solar north and west are indicated by the arrows labeled ''N'' and ''W''.
The field of view of the images is approximately 120'' $\times$ 150''.
The scan and the slit directions are also shown by the gray arrows.
The horizontal lines seen in the DST images are instrumental hair-line features along the scan direction.
(c) Time sequence images observed with SCIP in Ca~8542 line center.
The field of view of the images is approximately 54'' $\times$ 54''.
The white box indicates the field of view shown in panel (d).
(d) SCIP images obtained at 03:10:15~UT in different spectral lines.
From left to right, continuum, Fe~8514 line center, 
K~D$_1$ line center, and Ca~8498 line center are shown.
%
Solar north and west are indicated by the arrows.
The scan and the slit directions are shown by the gray arrows.
The time shown in each image indicates the start time of the corresponding raster scan.}
\label{fig:evolv}
\end{figure*}

The coordinated observations with SCIP and DST captured the temporal evolution of a compact flare kernel located within the sunspot umbra.
The time sequence of Ca~8542 line center images obtained with DST is shown in Figure~\ref{fig:evolv}~(a), and those obtained with SCIP in Figure~\ref{fig:evolv}~(c).
The kernel exhibits an apparent motion toward the umbral center (roughly in the eastward direction), approximately along a partial light bridge (penumbral intrusion into the umbra).
For clarity, the fields of view of both the DST and SCIP images are rotated by approximately 75$^{\circ}$ and 85$^{\circ}$, respectively, with respect to the solar north–south direction (see Fig.~\ref{fig:evolv}).
The apparent speed of the kernel is approximately 10~km~s$^{-1}$, estimated from the motion of its leading edge toward the umbra.

Here, we compare the appearance of the flare kernel in different spectral lines shortly before the flare peak.
Figure~\ref{fig:evolv}~(b) shows continuum (8537.45 -- 8537.97~{\AA}), Na~D$_1$ line center, and H$\alpha$ line center images obtained by DST.
Figure~\ref{fig:evolv}~(d) shows SCIP images in different spectral lines, continuum (8522.94 -- 8524.52~{\AA}), Fe~8514 line center, K~D$_1$ line center, and Ca~8498 line center.
Significant brightenings are detected in the chromospheric lines H$\alpha$, Ca~8542, and Na~D$_1$ in the DST observation, although the brightening is fainter in Na~D$_1$.
In the SCIP observations, clear brightenings are seen in Ca~8498, Ca~8542, whereas only a very slight enhancement is detected in K~D$_1$, as shown in the white arrow in Figure~\ref{fig:evolv}~(d). 
On the other hand, no clear brightening is detected in the photospheric lines.

\begin{figure*}[t]
\centering
\includegraphics[width=10cm]{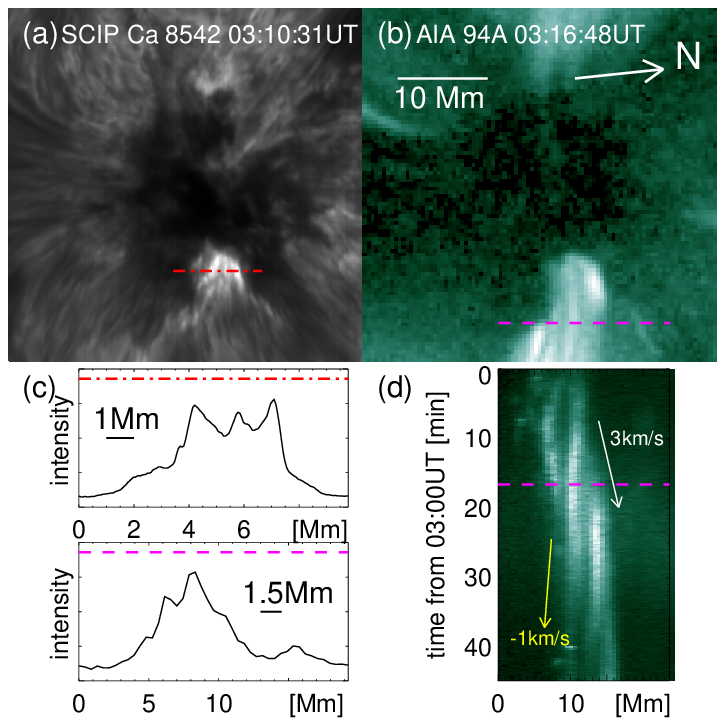}
\caption{
(a) SCIP Ca~8542 line center image taken in the raster scan starting at 03:10:31~UT.
The red dash-dotted line indicates the position across the flare kernel 
where the intensity profile shown in panel (c, top) was measured.
(b) SDO/AIA 94~{\AA} image taken at 03:16:48~UT.
The purple dashed line indicates the position used to derive the intensity
profiles shown in panel (c, bottom) and the time slice image shown in 
panel (d).
(c) {\it Top}: intensity profile across the flare kernel measured 
along the red line shown in panel (a).
{\it Bottom}: intensity profile across the flare loop measured 
along the purple line shown in panel (b).
The intensities are scaled arbitrarily.
(d) Time slice image constructed along the purple dashed line in panel (b),
showing the temporal evolution of fine structures in the flare loops.
}
\label{fig:slice}
\end{figure*}

The high spatial resolution observations with SCIP reveal that the flare kernel is composed of fine-scale substructures. 
The panels of Figure~\ref{fig:slice} (a) and (c) show the fine structures seen in the SCIP flare kernel.
The leading edge of the kernel is particularly bright.
In the trailing region, substructures appear as elongated, thread-like features arranged in parallel, with typical widths of about 1~Mm.
These substructures are more clearly identified in the relatively weaker trailing emission, and the apparent difference between the leading and the trailing regions should therefore be interpreted with caution.
It does not necessarily imply that the bright leading edge and the trailing substructures represent distinct evolutionary phases of the kernel.
Such internal structures are clearly resolved in the SCIP data owing to its higher spatial resolution, whereas they are not resolved in the DST data.

Structures with similar spatial scales are also seen in the associated flare loops.
As shown in Figure~\ref{fig:slice}~(c), fine-scale structures are visible within the flare loops in AIA 94~{\AA} images.
Although the chromospheric lines observed by SCIP and the coronal emission seen by AIA originate from different temperature regimes, the comparable spatial scales may suggest a connection between the chromospheric kernel substructures and the coronal flare loops.
The time slice image across the flare loops shown in Figure~\ref{fig:slice}~(d) reveals fine structures with spatial scales of approximately 1-1.5~Mm, consistent with those observed within the umbral flare kernel.
The loop system also exhibits systematic evolution.
In the time slice image, we can see that the loop bundle moves from left to right with a velocity of about 3~km~s$^{-1}$, as indicated with the white arrow.
This may reflect a combination of apparent motion due to ongoing magnetic reconnection and the temporal evolution of individual reconnected loops.
On the other hand, the fine magnetic field lines (threads) tend to shift from right to left, with a smaller velocity of about 1~km~s$^{-1}$ as shown with the yellow arrow.
This complex evolution of the flare loops suggests that the magnetic field is twisting or untwisting.

\section{Spectral Characteristics}

\begin{figure*}[t]
\centering
\includegraphics[width=10cm]{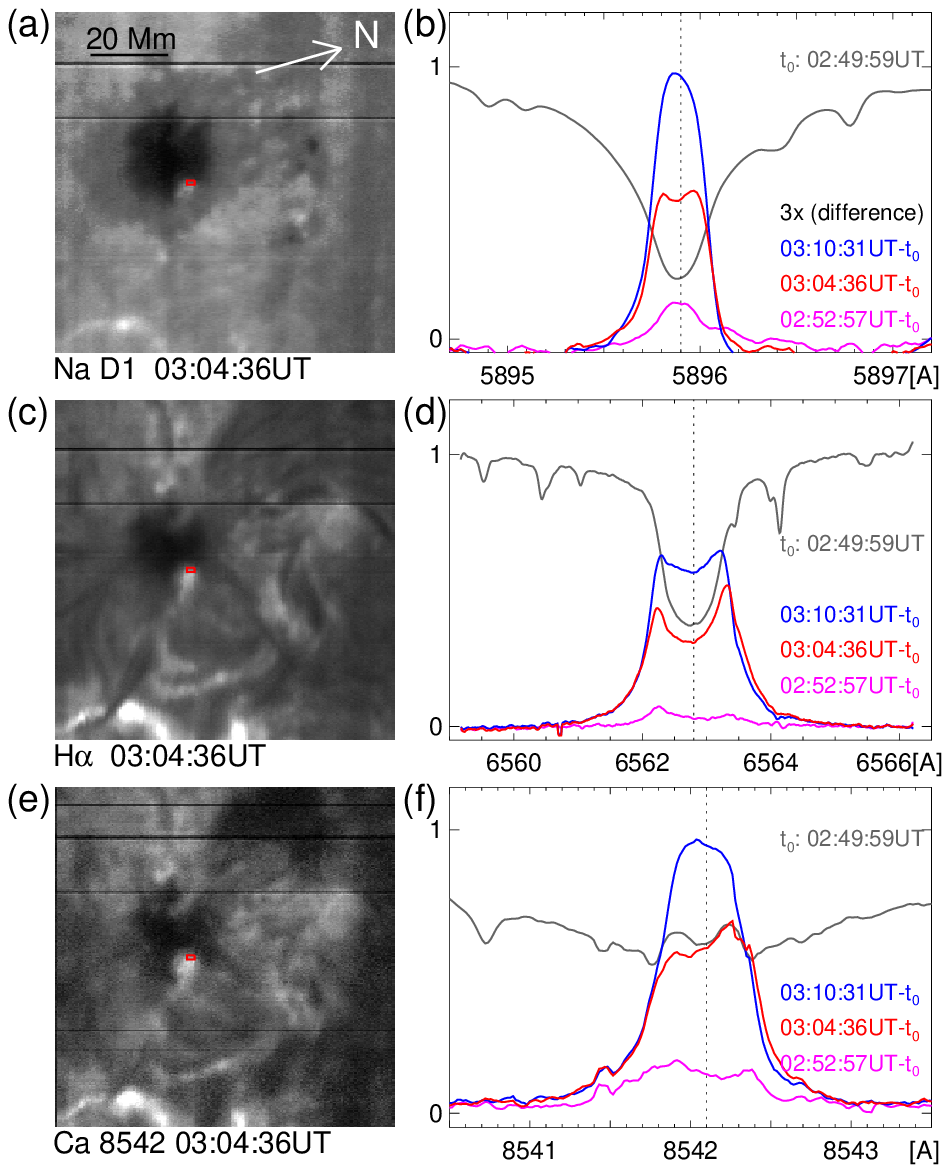}
\caption{(a) Line center image of the Na~D$_1$ at 03:04:36~UT obtained with DST.
The red box indicates the flare kernel region from which the spectra shown in panel (b) were averaged.
(b) Temporal evolution of the Na~D$_1$ spectra averaged over the region indicated in panel (a).
The spectrum at 02:49:59~UT is shown in gray and used as a reference ($t_0$).
Colored curves represent difference spectra relative to $t_0$ at 02:52:57~UT (purple), 03:04:36~UT (red), and 03:10:31~UT (blue), respectively.
(c, d) Same as panels (a) and (b), respectively, but for the H$\alpha$ line.
(e, f) Same as panels (a) and (b), respectively, but for the Ca~8542 line.
The difference spectra are multiplied by a factor of 3 in panel (b) for clarity.
All spectra in panels (b), (d), and (f) are normalized by the local continuum level and are smoothed using a 5-point running average.
The time shown for each raster image or spectrum indicates the start time of the corresponding raster scan.}
\label{fig:DST}
\end{figure*}

Simultaneous imaging spectroscopy in multiple chromospheric lines enables us to investigate the temporal evolution and spectral characteristics of flare kernels in the umbra.
Figure~\ref{fig:DST} shows the temporal evolution of the flare kernel spectra observed with DST in multiple chromospheric lines.
A pre-flare spectrum at the same spatial region is used as the reference, as shown by the gray curves.
The apparent double-peaked structure seen in the difference spectra of many chromospheric lines mainly reflects the reduced intensity near the optically thick line core relative to the wings.
In this paper, red asymmetry refers to an excess of emission toward the red side of the line profile, irrespective of the apparent double-peaked structure of the difference spectra.
At 03:04:36~UT, corresponding to the early stage of kernel formation before the flare peak, the spectra exhibit a clear enhancement on the red side, resulting in a red asymmetry.
At an earlier time (02:52:57~UT), the spectral signal is much weaker, and a slight excess on the blue side is observed instead.
At a later time (03:10:31~UT), when the kernel intensity increases, the line profiles become more symmetric, and the red asymmetry is less pronounced.
These observations indicate that the spectral asymmetry is most pronounced during the initial phase of kernel formation and weakens as the kernel brightening develops.

\begin{figure*}[t]
\centering
\includegraphics[width=10cm]{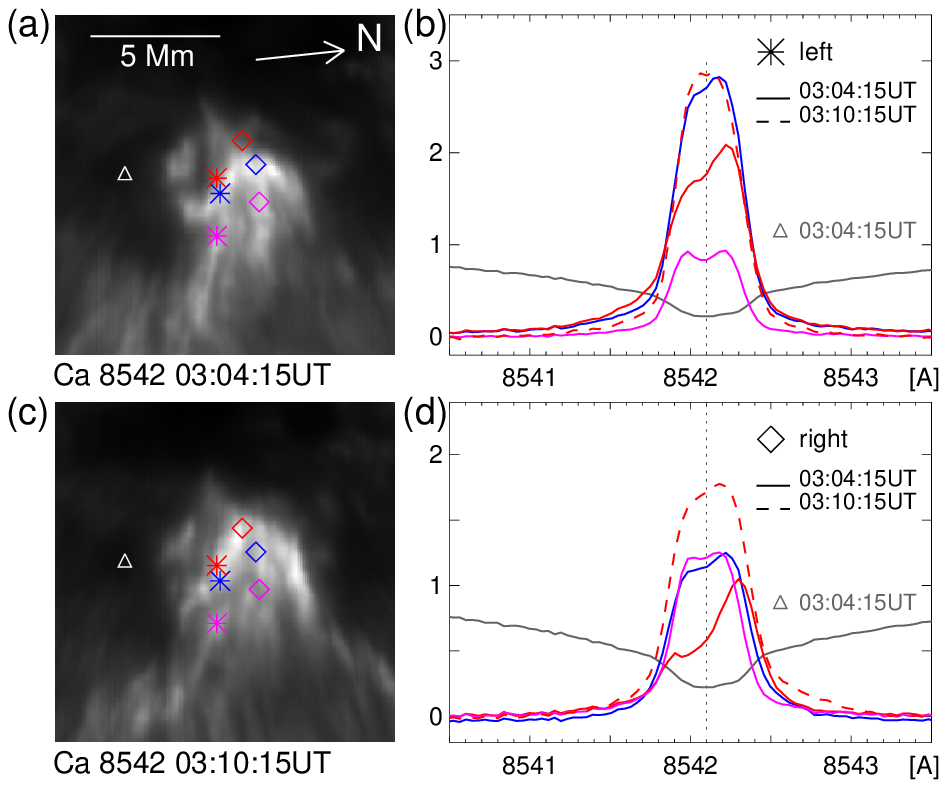}
\caption{
(a) Ca~8542 line-center image of the flare kernel at 03:04:15~UT observed with SCIP.
Colored symbols indicate the locations within the flare kernel from which the spectra shown in panel (b) were extracted.
The asterisks $\ast$ are located along the left thread, while the diamonds $\diamond$ mark the right thread.
The triangle $\triangle$ indicates a nearby reference point in the umbra (outside the kernel).
(b) Ca~8542 spectra obtained at the positions along the left thread indicated by the asterisk $\ast$ in panel (a).
Solid and dashed curves represent spectra at 03:04:15~UT and 03:10:15~UT, respectively.
(c) Same as panel (a), but for observations at 03:10:15~UT.
(d) Same as panel (b), but for the right thread.
In panels (b) and (d), all spectra are normalized to the local continuum level.
The spectra at the reference point indicated by triangle $\triangle$ in panels (a) and (c) are shown in gray.
The other spectra are shown as difference spectra relative to the reference point.
The time shown for each raster image or spectrum indicates the start time of the corresponding raster scan.}
\label{fig:SCIP}
\end{figure*}

The high spatial resolution of SCIP reveals the internal structure of the flare kernel.
These fine structures exhibit different spectral characteristics depending on their location.
Figure~\ref{fig:SCIP} presents Ca~8542 spectra obtained at multiple locations within the flare kernel using SCIP.
A nearby umbral spectrum taken at the same time is used as the reference, as shown by the gray curves.
At given locations (the red asterisk $\ast$ and diamond $\diamond$ points), the spectra show a temporal evolution similar to that seen in Figure~\ref{fig:DST}.
That is, by comparing the red solid and red dashed lines in Figure~\ref{fig:SCIP}(b) and (d), we can observe a transition from red-asymmetric profiles to more symmetric ones as the kernel brightens.

Comparisons among different positions along a thread within the kernel at the same time reveal variations in spectral asymmetry and intensity, as illustrated by the red, blue, and purple solid lines in Figure~\ref{fig:SCIP}(b) and (d).
When the apparent motion of the kernel associated with the progress of magnetic reconnection is taken into account, these spatial differences can also be interpreted as reflecting different stages of the temporal evolution of the kernel.
In this sense, the spectra depicted by the red lines show stronger redward asymmetry, whereas those labeled in blue and purple exhibit weaker asymmetry. 
We also note that some early-phase profiles show a blue-wing enhancement in addition to the redward asymmetry, as also suggested by the slight blue-side excess in the early DST spectra (Figure~\ref{fig:DST}). 
Similar tendencies are also seen at the other sampled locations, although the absolute intensities and asymmetry strengths differ.

\section{Discussion} 

In this study, we have investigated a compact flare kernel located in a sunspot umbra during the M1.4 flare on 2024 July 13, using coordinated imaging spectroscopic observations with SCIP and DST.
Simultaneous observations in multiple chromospheric lines enabled us to examine both the temporal development and spatial structure of the flare kernel with high spatial and spectral resolution.
High spatial resolution observations with SCIP reveal that the umbral flare kernel is composed of multiple fine-scale substructures with characteristic spatial scales of approximately 1000~km.
The observed spatial scales are comparable to the fine structures seen in associated flare loops in AIA 94~{\AA} images.
Spectroscopic observations with DST detect red-shifted components (red asymmetry) in multiple chromospheric lines during the early phase of flare kernel development, consistent with chromospheric condensation.
With the higher spatial resolution, SCIP reveals that individual fine kernels exhibit red-shifted profiles.
In the SCIP Ca~8542 sequence, strong red-asymmetric profiles at individual fine kernels are typically visible only over a small number of raster scans, suggesting a timescale comparable to the raster cadence of about 40 seconds.
This is qualitatively consistent with previous IRIS observations, which found red-shifted chromospheric components evolving on timescales of several tens of seconds, typically about 30 seconds \citep{Graham2015,Graham2020}.
These results demonstrate the importance of high-resolution, multi-line imaging spectroscopy for resolving the fine-scale dynamics of flare energy deposition in the chromosphere, particularly within sunspot umbrae.

Previous studies \citep[e.g.,][]{Shoji1995} reported differences in redshifts among chromospheric lines and interpreted them as a consequence of the formation-height dependence of chromospheric condensation, with stronger downward motions occurring in the upper chromosphere.
The DST observations in this study show a similar tendency.
At the early phase of the flare kernel development (03:04:36~UT), pronounced red asymmetry is detected in H$\alpha$ and Ca~8542, whereas the Na~D$_1$ line exhibits only a modest red shift (Fig.~\ref{fig:DST}).
Although the spectral profiles in the present study are more complex, the relative differences among these chromospheric lines remain consistent with this interpretation.
\citet{Kuckein2025}, on the other hand, emphasized that the formation height and velocity diagnostics of Ca~8542 in flare kernels can be uncertain, and demonstrated the importance of simultaneous He~{\sc i}~10830~{\AA} observations for constraining the chromospheric response.

The high spatial resolution of SCIP enables us to resolve the fine internal structure of the flare kernel.
SCIP Ca~8542 images reveal that flare kernels are composed of multiple fine-scale substructures.
Each fine kernel shows enhanced emission and red-shifted profiles only for a limited period, indicating that energy deposition and subsequent chromospheric condensation occur intermittently and locally.
In this study, we demonstrate that what has appeared as a single flare kernel at lower spatial resolution is in fact composed of multiple fine-scale substructures.
A simple spatial degradation of the SCIP Ca~8542 image to a scale comparable to the DST resolution also makes these fine substructures much less distinct.
This suggests that the red-shifted signatures reported in \citet{Ichimoto1984} do not necessarily imply a continuous downward motion within a single flare kernel.
Instead, the apparent persistence of red asymmetry can be understood as a temporal succession of multiple fine kernels undergoing localized condensation.
The similar temporal evolution seen at different locations within the kernel supports the idea that the apparent persistence of red asymmetry reflects successive energy release in multiple fine kernels.
The apparent motion of the kernel and the spatial variation of the spectral profiles imply that the location of strong energy deposition changes with time.
This spatially resolved view extends the temporal picture proposed by \citet{Ichimoto1984}, indicating that the formation and evolution of flare kernels are governed by temporally intermittent energy release in multiple fine-scale structures.

\begin{acknowledgments}
The authors thank the anonymous referee for constructive comments that significantly improved the quality of this paper.
\textsc{Sunrise~iii} is supported by funding from the Max-Planck-F\"{o}rderstiftung (Max Planck Foundation), NASA under Grants \#80NSSC18K0934 and \#80NSSC24M0024 (“Heliophysics Low Cost Access to Space” program), and the ISAS/JAXA Small Mission-of-Opportunity program and JSPS KAKENHI Grant Numbers JP18H05234 (PI: Y. Katsukawa) and JP23K25916 (PI: Y. Katsukawa). 
This research has received financial support from the European Union’s Horizon 2020 research and innovation program under grant agreement No. 824135 (SOLARNET) and No. 101097844 (WINSUN) from the European Research Council (ERC). 
It has also been funded by the Deutsches Zentrum f\"{u}r Luft- und Raumfahrt e.V. (DLR, grant no. 50 OO 1608). 
The Spanish contributions have been funded by the Spanish MCIN/AEI under projects RTI2018-096886-B-C5, PID2021-125325OB-C5, and PID2024-156066OB-C5, and from “Center of Excellence Severo Ochoa” awards to IAA-CSIC (SEV-2017-0709, CEX2021-001131-S), all co-funded by European REDEF funds, "A way of making Europe”.
We also express our sincere gratitude to the staff of the Hida Observatory for developing and maintaining the instruments and daily observation. 
We would like to acknowledge the data use from GOES. 
This work was supported by JSPS KAKENHI Grant Numbers JP24K07093 (PI: A. Asai), JP23KJ0299 (PI: R.T. Ishikawa), JP24K07105 (PI: M. Kubo), JP23K13152 (PI: Y. Kawabata), and JP21K13972 (PI: T. Oba).
Parts of the English editing and manuscript preparation were assisted by using ChatGPT (OpenAI).
\end{acknowledgments}

\begin{contribution}
All authors contributed equally to the \textsc{Sunrise~iii}/SCIP and Hida/DST collaboration.
\end{contribution}

\facilities{\textsc{Sunrise~iii}~(SCIP), Hida~(DST)}

\software{IDL-SolarSoftWare} 

\bibliography{SCIP-DSTpap}{}
\bibliographystyle{aasjournalv7}

\end{document}